\RequirePackage[running]{lineno}

\documentclass[aps,prl,nofootinbib,twocolumn,showpacs,superscriptaddress,letterpaper,amsmath,amssymb]{revtex4}

\usepackage{graphicx}  % needed for figures
\usepackage{dcolumn}   % needed for some tables
\usepackage{bm}        % for math
\usepackage{amssymb}   % for math
\usepackage{amsmath}
\usepackage{feynmf}%%%{feynmp}
\usepackage{slashed}
\usepackage{multirow}
\def\gsim{\lower0.5ex\hbox{$\:\buildrel >\over\sim\:$}}
\def\lsim{\lower0.5ex\hbox{$\:\buildrel <\over\sim\:$}}

\newcommand{\be}{\begin{equation}}
\newcommand{\ee}{\end{equation}}
\newcommand{\bea}{\begin{eqnarray}}
\newcommand{\eea}{\end{eqnarray}}

\newcommand{\nbox}{{\,\lower0.9pt\vbox{\hrule \hbox{\vrule height 0.2 cm
\hskip 0.2 cm \vrule height 0.2 cm}\hrule}\,}}

\begin{document}

\thispagestyle{empty}
\vspace*{-3.5cm}

\vspace{0.5in}

%\begin{flushright}
%\today\\
%\end{flushright}
%\vspace{0.5in}
%\title{Missing the WIMP Forest for the $Z$'s}
\title{Sensitivity of potential future $pp$ colliders to quark compositeness}

\begin{center}
\begin{abstract}
A study is presented of the sensitivity of potential future $pp$
colliders to quark compositeness.  The analysis uses normalized dijet
angular distributions compared to expectations from leading-order
contact interaction models.
\end{abstract}
\end{center}

\author{Leonard Apanasevich}
\affiliation{Department of Physics, University of Illinois at Chicago,
  Chicago, IL, 60607}
\author{Suneet Upadhyay}
\affiliation{Department of Physics and Astronomy, University of
  California, Irvine, CA 92697}
\author{Nikos Varelas}
\affiliation{Department of Physics, University of Illinois at Chicago,
  Chicago, IL, 60607}
\author{Daniel Whiteson}
\affiliation{Department of Physics and Astronomy, University of
  California, Irvine, CA 92697}
\author{Felix Yu}
\affiliation{Fermi National Accelerator Laboratory, Batavia, IL 60510}

%\preprint{UCI-HEP-TR-2012-XX}
\maketitle

% introduction
%\linenumbers

%\subsection{Introduction}

Though quarks are treated as fundamental particles in the standard
model, facilities with sufficient center-of-mass energy may reveal
substructure~\cite{Eichten:1983hw,Eichten:1984eu, Chiappetta:1990jd,
  Lane:1996gr}. Quarks as bound states of more fundamental particles
may explain current outstanding questions, such as the number of quark
generations, the charges of the quarks, or the symmetry between the
quark and lepton sectors.

\begin{figure}
\includegraphics[width=0.95\linewidth]{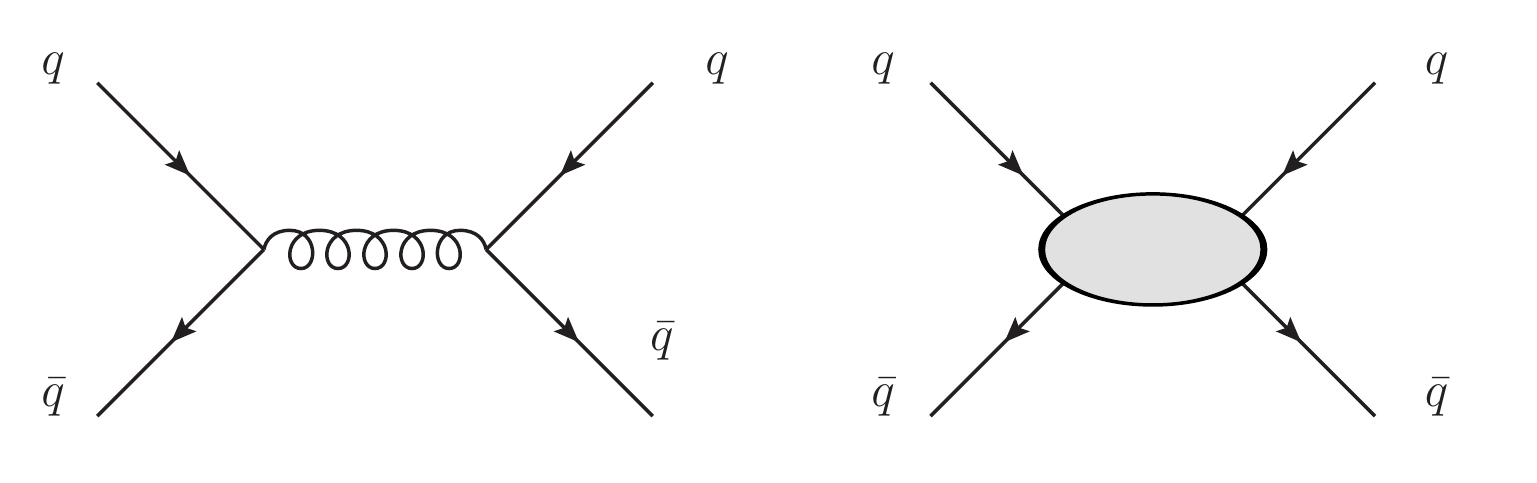}
\caption{Diagrams for QCD mediation of quark-quark interactions (left)
  and a four-fermion contact interaction (right) describing an
  effective field theory for the mediation of a new interaction
  between quark constituents.}
\label{fig:diag}
\end{figure}

A typical approach to the study of quark
compositeness~\cite{Chatrchyan:2012bf} is to search for evidence of
new interactions between quarks at a large characteristic energy
scale, $\Lambda$.  At interaction energies below $\Lambda$, the
details of the new interaction and potential mediating particles can
be integrated out to form a four-fermion contact interaction model
(see Fig.~\ref{fig:diag}).

 This is well-described by an
effective field theory approach~\cite{Beringer:1900zz}:

\begin{align} 
L_{qq} =  \frac{2\pi}{\Lambda^2}&[\eta_{LL}(\bar{q_L}\gamma^\mu
q_L)(\bar{q_L}\gamma^\mu q_L)\\\nonumber
&  + \eta_{RR}(\bar{q_R}\gamma^\mu q_R)(\bar{q_R}\gamma^\mu q_R) \\\nonumber
&  + 2\eta_{RL}(\bar{q_R}\gamma^\mu q_R)(\bar{q_L}\gamma^\mu q_L) ]
\end{align}

\noindent
where the quark fields have $L$ and $R$ chiral projections and the
coefficients $\eta_{LL}$, $\eta_{RR}$, and $\eta_{RL}$ turn on and off
various interactions. In this study, we examine the cases of energy
scales $\Lambda^+_{LL}$, $\Lambda^+_{RR}$, and $\Lambda^+_{V-A}$ with
couplings $(\eta_{LL},\eta_{RR}, \eta_{RL}) = (1,0,0), (0,1,0)$ and
$(0,0,1)$, respectively, in order to demonstrate the center-of-mass
dependence of the sensitivity of possible future $pp$ facilities.

Evidence for contact interactions would appear in dijets with large
values of dijet mass ($m_{jj}$) and small values of $\theta^*$, the
center-of-mass angle relative to the beam axis. Dijets produced via
quantum chromodynamics (QCD) are predominantly at small values of
$m_{jj}$ and large values of $\theta^*$, as they are peaked in the
forward and backward directions.

\begin{figure}
\includegraphics[width=0.45\linewidth]{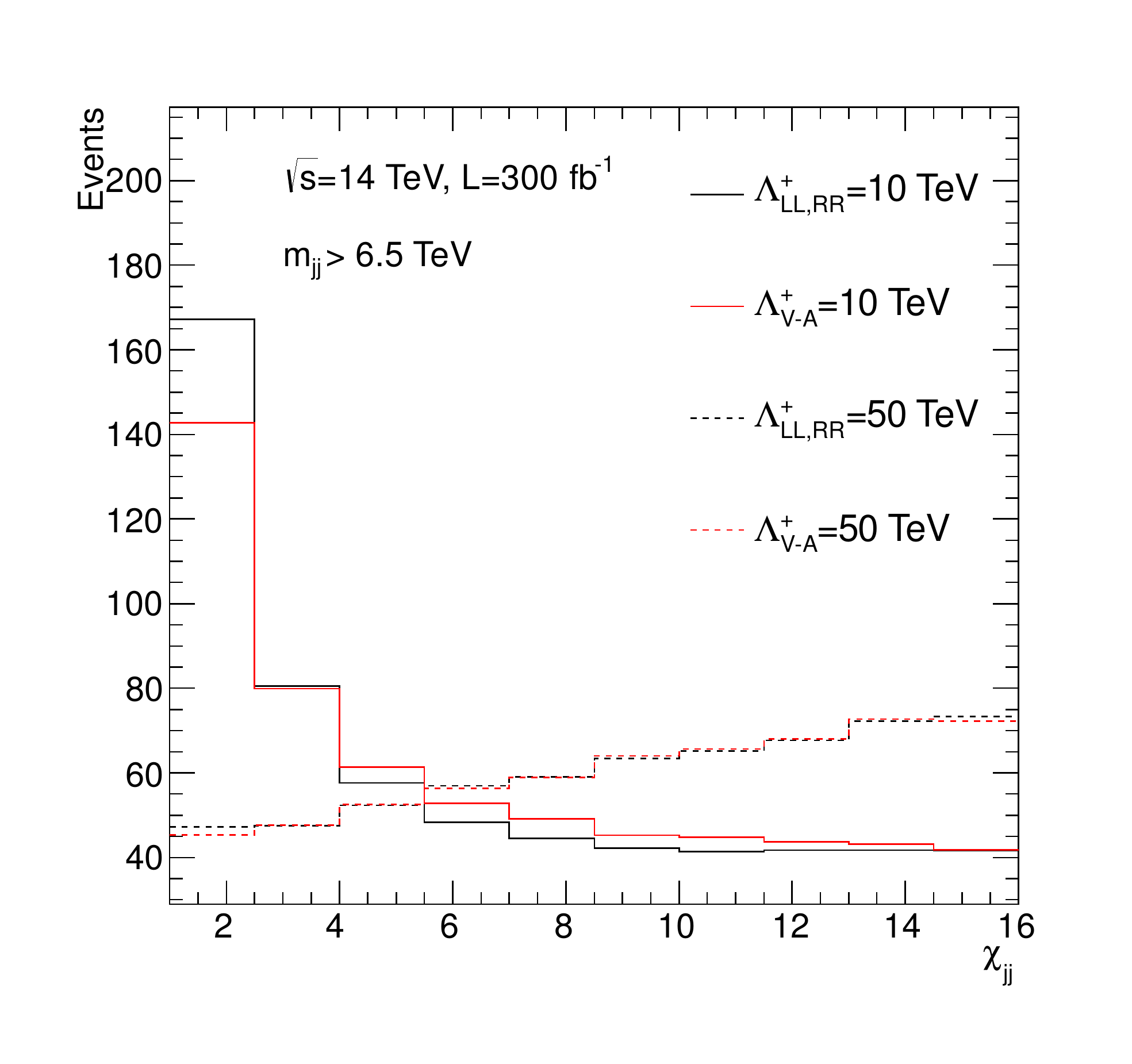}
\includegraphics[width=0.45\linewidth]{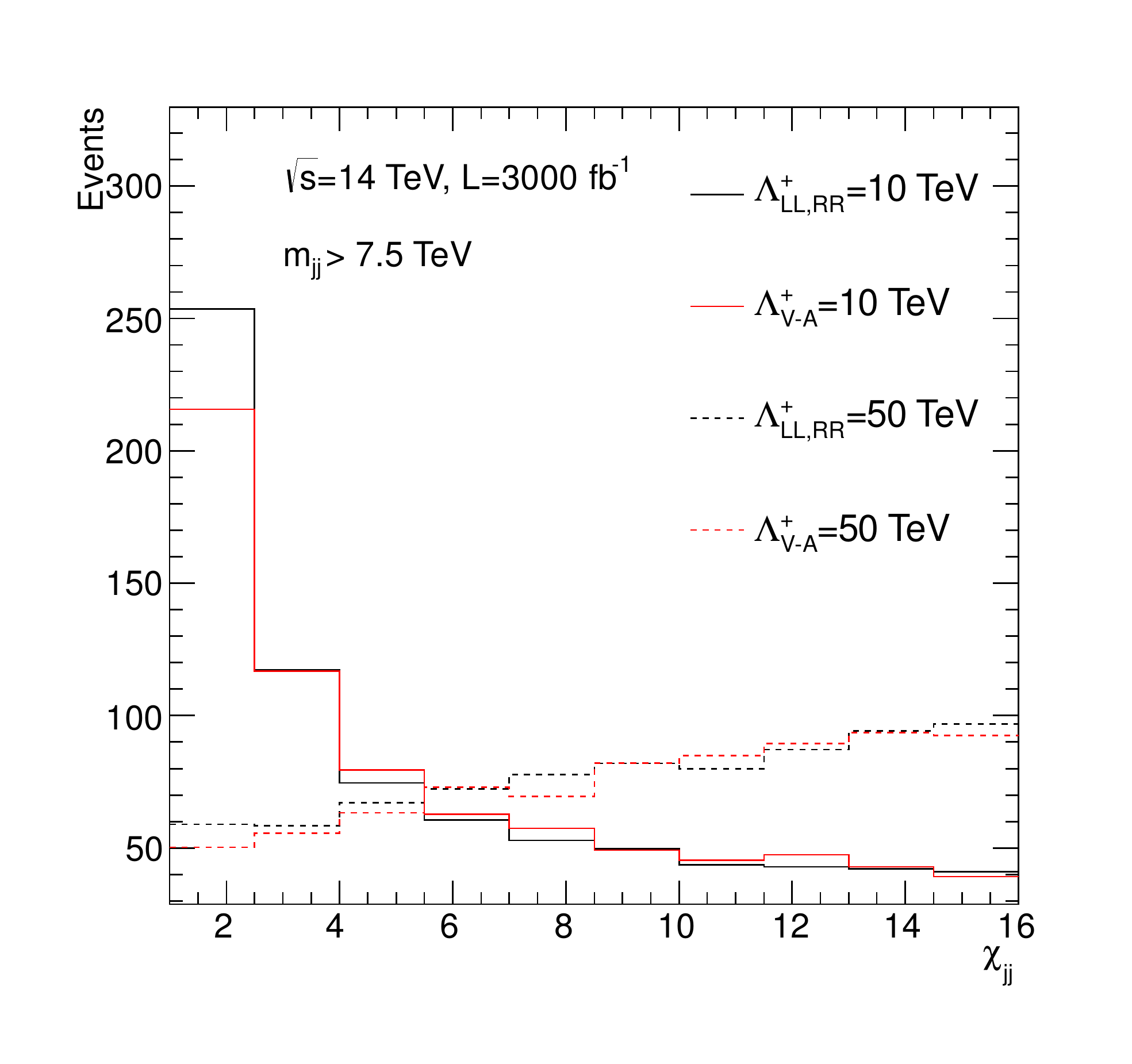}
\includegraphics[width=0.45\linewidth]{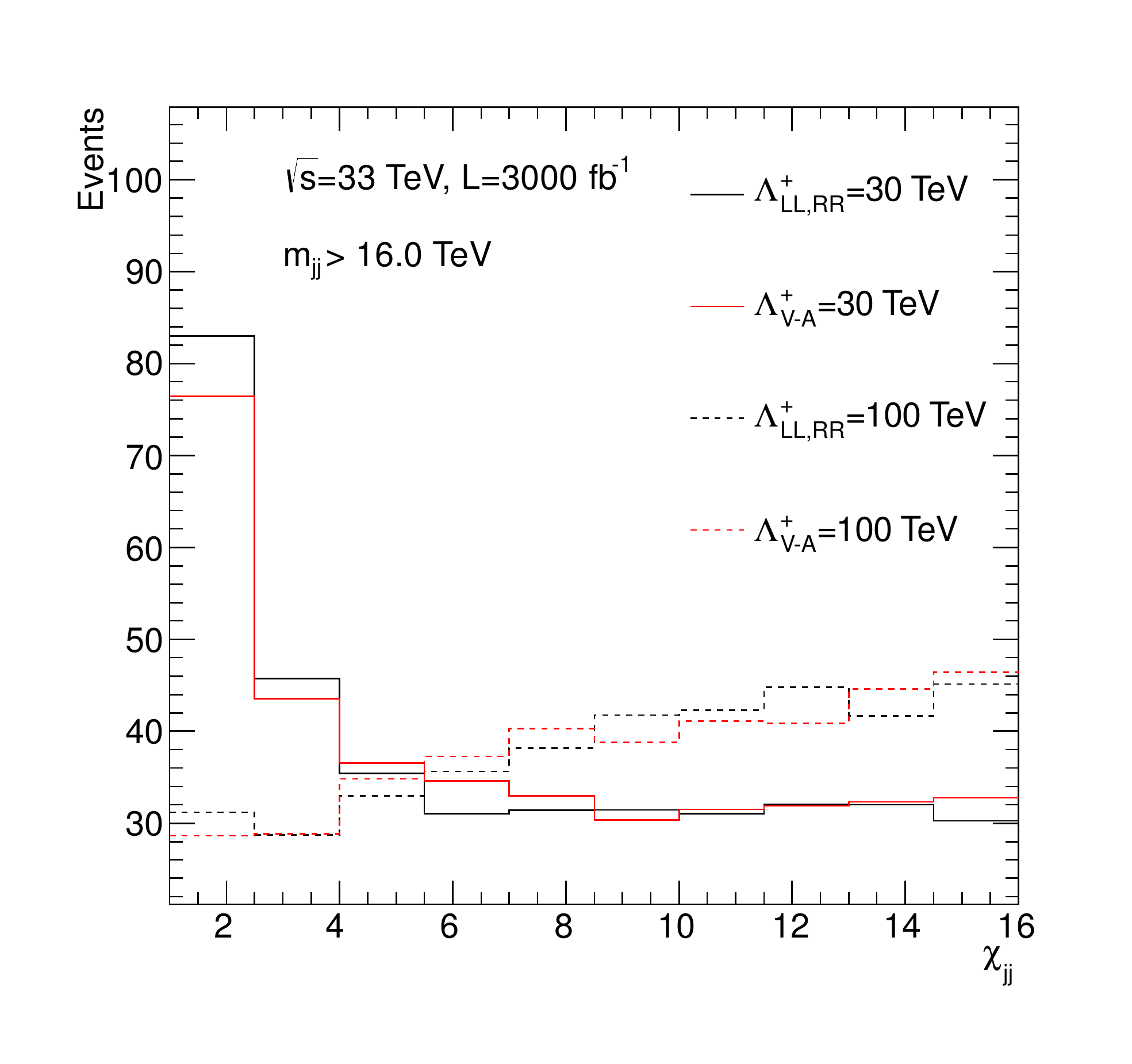}
\includegraphics[width=0.45\linewidth]{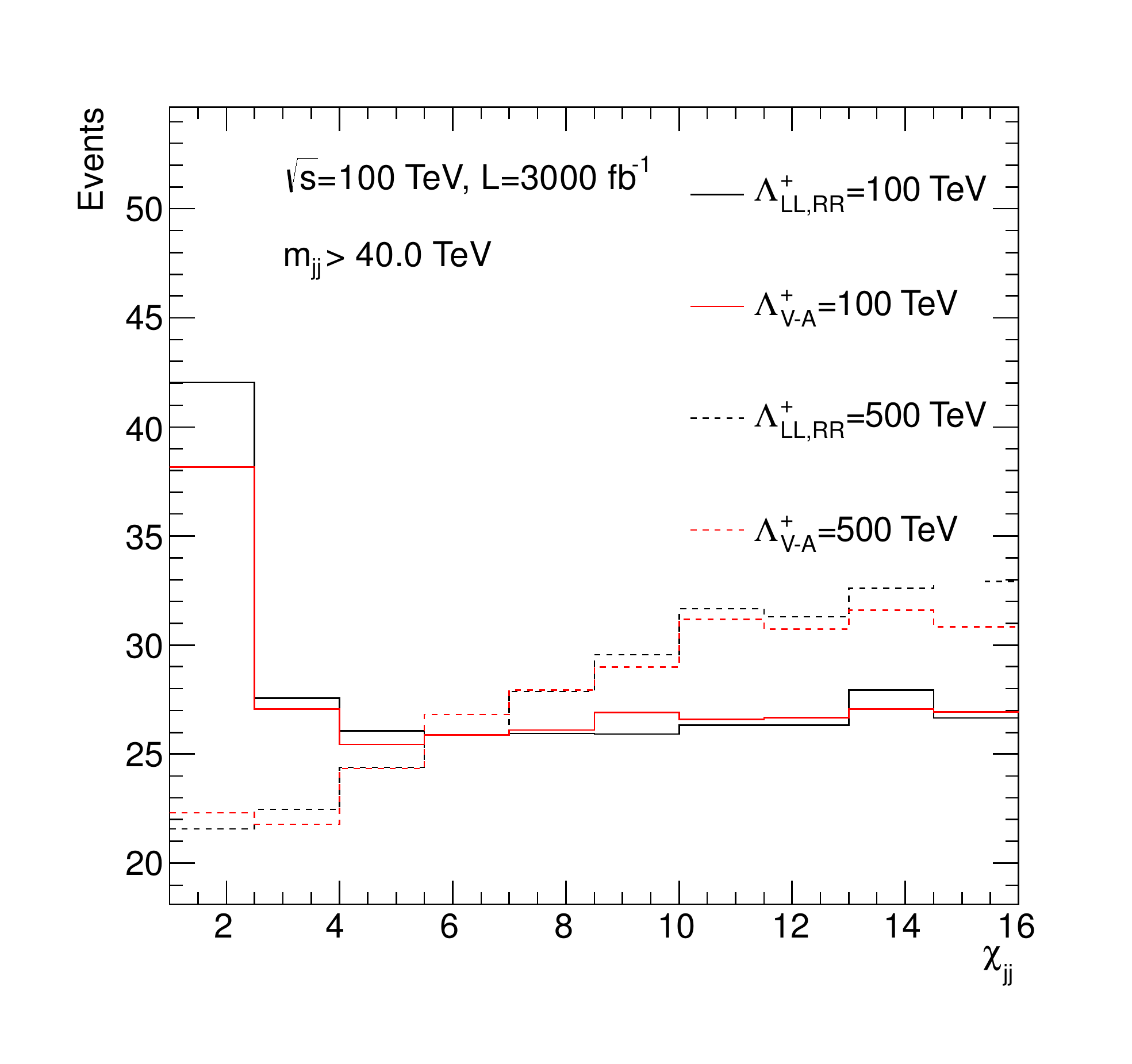}

\caption{ Distributions of $\chi_{jj}$ for QCD and contact
  interactions with two choices of $\Lambda^+_{LL,RR}$, and
  $\Lambda^+_{V-A}$, for the four collider facilities considered.}
\label{fig:coup}
\end{figure}

Dijet angular distributions can also be used to search for a variety
of new physics models, such as those hypothesized to mediate the interaction between the standard model
sector and the dark matter sector~\cite{Dreiner:2013vla}, and extra spatial dimensions~\cite{ArkaniHamed:1998rs, Cheung:2001mq,
  Pomarol:1998sd, Dienes:1998vg, Atwood:1999qd}.

In this study, we generate events at leading-order with {\sc
  madgraph}~\cite{Alwall:2011uj}, describe the showering and
hadronization with {\sc pythia}~\cite{Sjostrand:2007gs} and the
detector response with {\sc delphes}~\cite{Ovyn:2009tx} for the
facilities described in Table~\ref{tab:fac}.

Following the approach of Ref.~\cite{Chatrchyan:2012bf}, we calculate
$\chi_{jj} = e^{|y_1 - y_2|}$ where $y_1$ and $y_2$ are the rapidities
of the two highest transverse momentum (leading) jets.  The
distribution for QCD interactions is slightly increasing with
$\chi_{jj}$, while contact interaction models predict angular
distributions that are strongly peaked at low values of $\chi_{jj}$,
 as can be seen in Fig.~\ref{fig:coup}. Note that $LL$
and $RR$ models have no significant differences in their predicted
distributions of $\chi_{jj}$. In this study, $|y_1|$ and $|y_2|$ are
restricted to be less than 2.5 and $\chi_{jj}$ less than 16.

\begin{figure}
\includegraphics[width=0.9\linewidth]{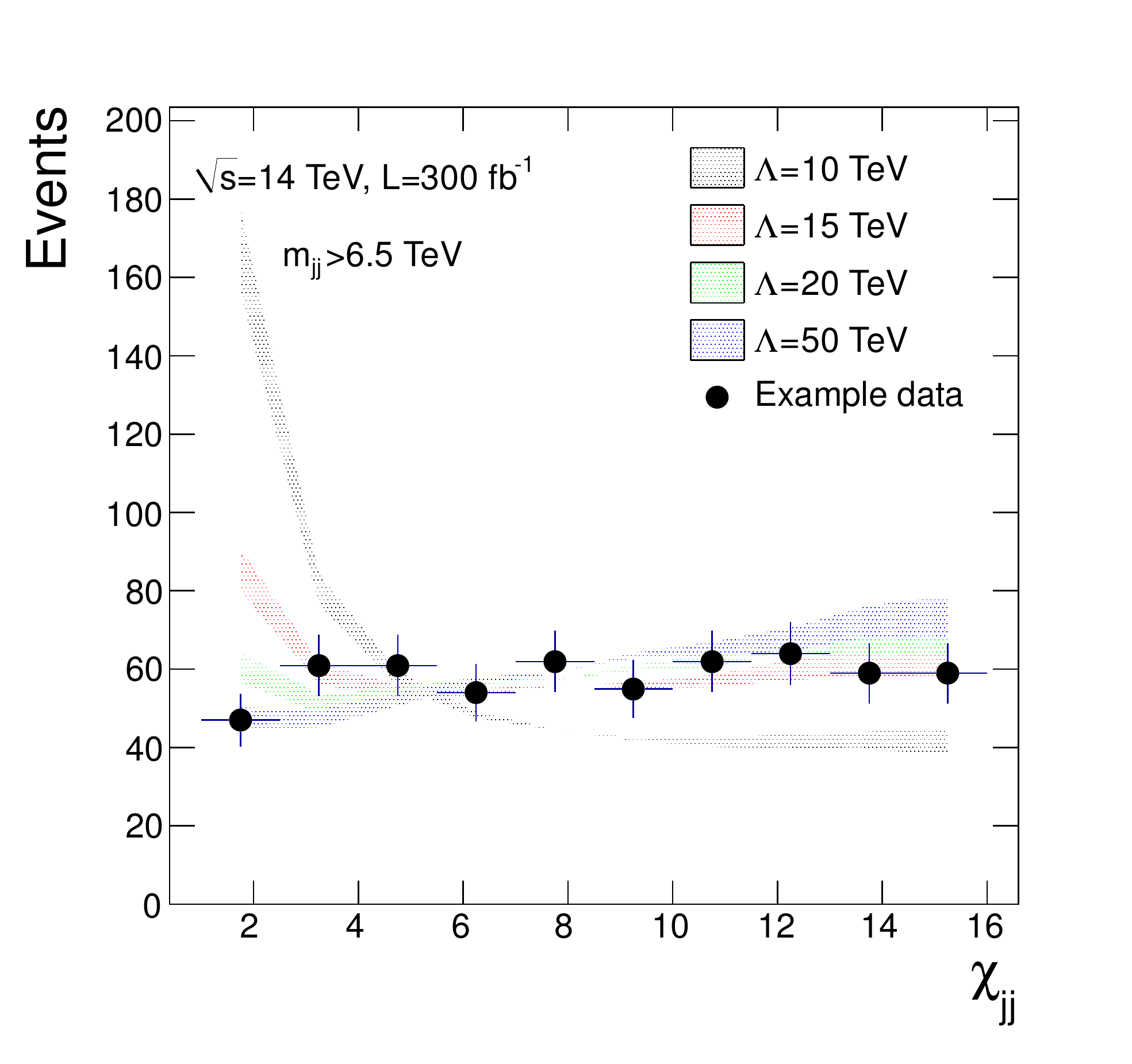}
\caption{ Distributions of $\chi_{jj}$ for QCD and contact
  interactions with a variety of choices of $\Lambda_{LL,RR}$ for the
  case of $pp$ interactions with $\sqrt{s}=14$~TeV and
  $\mathcal{L}=300$~fb$^{-1}$.}
\label{fig:dijet_14tev300}
\end{figure}

\begin{figure}
\includegraphics[width=0.9\linewidth]{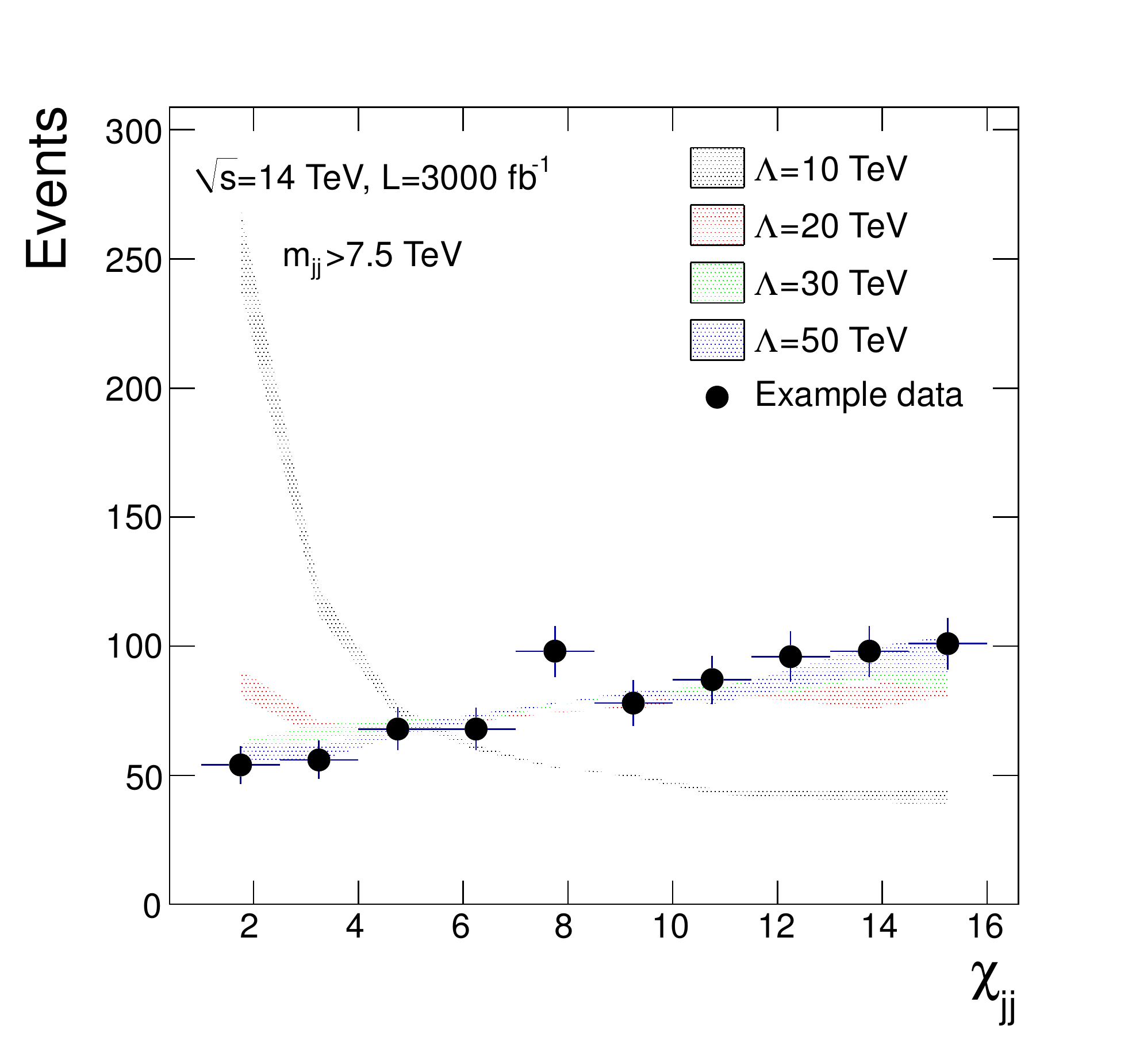}
\caption{Distributions of $\chi_{jj}$ for QCD and contact interactions
  with a variety of choices of $\Lambda_{LL,RR}$, for the case of $pp$
  interactions with $\sqrt{s}=14$~TeV and
  $\mathcal{L}=3000$~fb$^{-1}$.}
\label{fig:dijet_14tev3000}
\end{figure}

In the analysis of Ref.~\cite{Chatrchyan:2012bf}, the dominant
uncertainties are statistical and theoretical, followed by
experimental uncertainties such as jet energy resolution and
calibration.  Theoretical uncertainties are due primarily to variation
in the predicted $\chi_{jj}$ shape with changes to the renormalization
and factorization scales.  We extract the approximate size and
dependence on $\chi_{jj}$ of the theoretical uncertainty from
Ref.~\cite{Chatrchyan:2012bf} and apply it to the predictions from our
simulated samples.

The statistical analysis is performed by evaluating the likelihood
ratio
\[ q = -2 \ln
\frac{L(\Lambda + \textrm{QCD})}{L(\textrm{QCD})} \]
\noindent
where nuisance parameters are fixed at the nominal values.  The
extraction of the limit is done using the CLs~\cite{cls1,cls2}
technique, where the null and alternate hypothesis $p$-values are
evaluated from distributions of $q$ constructed using simulated
experiments in which the nuisance parameters have been varied
according to their prior probabilites.

The distortion of the $\chi_{jj}$ shape from new physics compared to
QCD is most distinct at large $m_{jj}$: however, the cross section
falls sharply with $m_{jj}$, reducing the statistical power of the
data.  These two effects are in tension, and there is an optimum lower
threshold value of $m_{jj}$ which captures the distortions without
sacrificing statistical power.  Note that in
Ref.~\cite{Chatrchyan:2012bf} the analysis is done in the two highest
bins of $m_{jj}$ and the lower bins are used to validate the QCD
predictions. In our study, we use only the highest bin in $m_{jj}$,
with modest loss of sensitivity.  We are able to reproduce the
sensitivity of Ref.~\cite{Chatrchyan:2012bf} using our approximate
approach.

\begin{table}
\caption{Details of compositeness studies for current and potential
  future $pp$ colliders, including center-of-mass energy ($\sqrt{s}$),
  total integrated luminosity ($\mathcal{L}$) and the minimum
  threshold in $m_{jj}$.}
\label{tab:fac}
\begin{tabular}{lrrrrrr}
\hline\hline
$\sqrt{s}$  & $\mathcal{L}$  &\ \ \  $m_{jj}$ threshold\ \ \ & $p_{T}$
threshod \\
(TeV) & (fb$^{-1}$) &  (TeV) & (TeV) \\
\hline
7 & 2 & 3 & 0.3 \\
14 & 300 & 5 & 1 \\
14 & 3000 & 7.5 & 1\\
33 & 3000 & 16 & 2.5\\
100 & 3000 & 44 & 5 \\
\hline\hline
\end{tabular}
\end{table}

\begin{figure}
\includegraphics[width=0.9\linewidth]{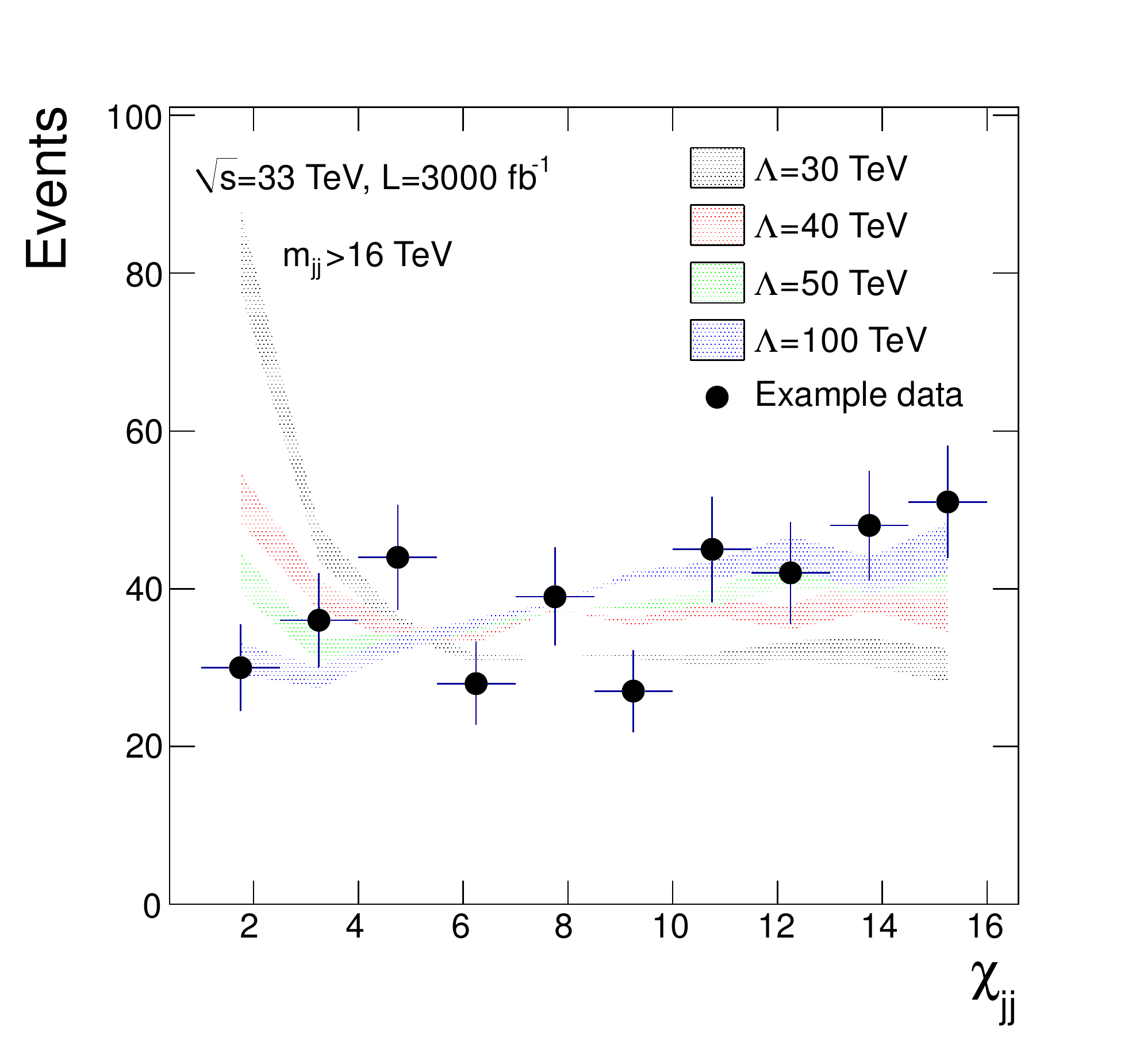}
\caption{ Distributions of $\chi_{jj}$ for QCD and contact
  interactions with a variety of choices of $\Lambda_{LL,RR}$ for the
  case of $pp$ interactions with with $\sqrt{s}=33$~TeV and
  $\mathcal{L}=3000$~fb$^{-1}$.}
\label{fig:dijet_33tev3000}
\end{figure}

\begin{figure}
\includegraphics[width=0.9\linewidth]{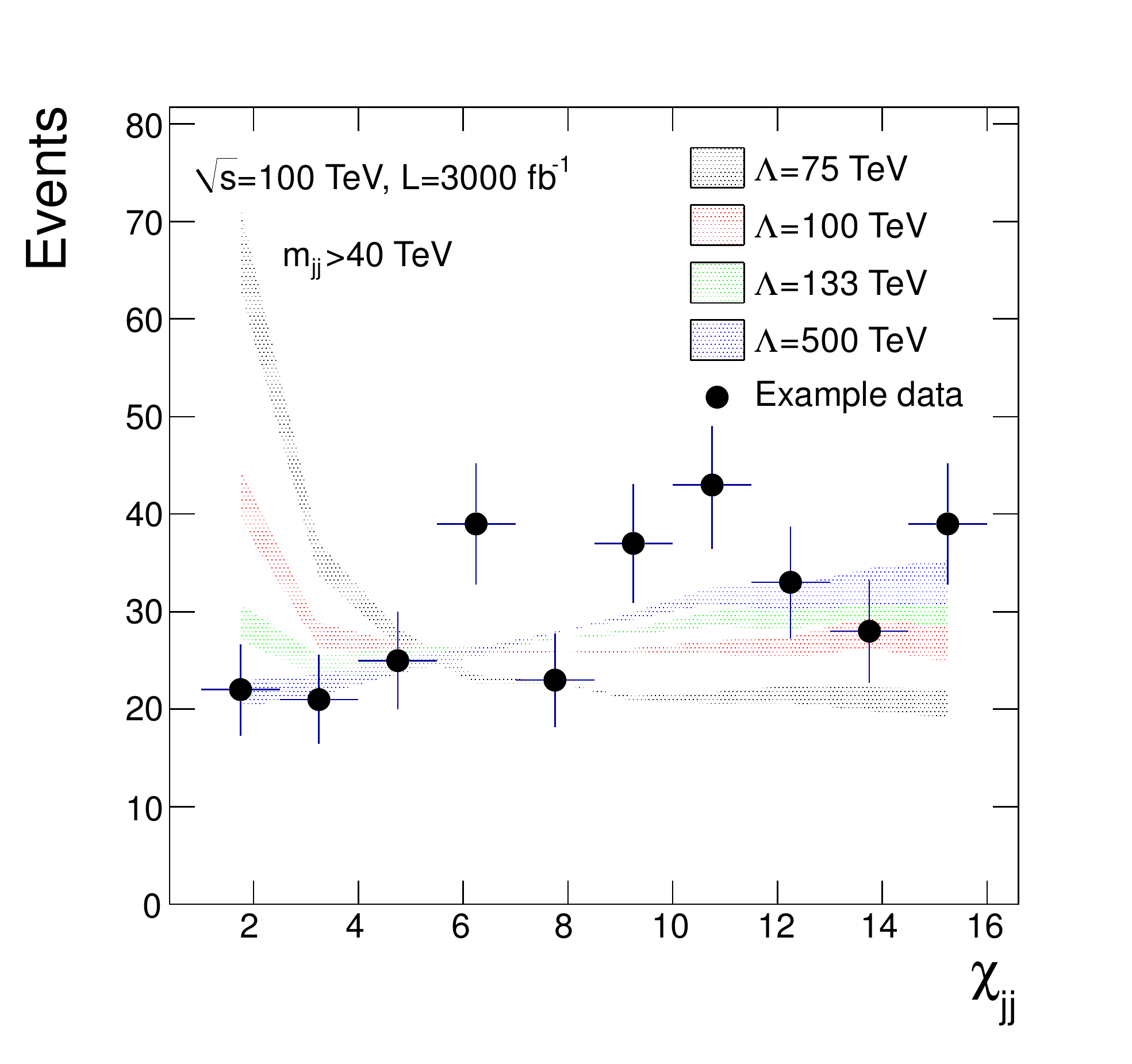}
\caption{ Distributions of $\chi_{jj}$ for QCD and contact
  interactions with a variety of choices of $\Lambda_{LL,RR}$. for the
  case of $pp$ interactions with with $\sqrt{s}=100$~TeV and
  $\mathcal{L}=3000$~fb$^{-1}$.}
\label{fig:dijet_100tev3000}
\end{figure}

If a deviation from QCD production is seen at the LHC with
$\sqrt{s}=14$ TeV, then a future facility may have the energy to
directly probe the new physics process.  If the deviation is due to
quark substructure, then with sufficiently high energy and luminosity
it is possible to observe the quark constituents and investigate their
interactions.

\subsection{Conclusions}

\begin{figure}
\includegraphics[width=0.9\linewidth]{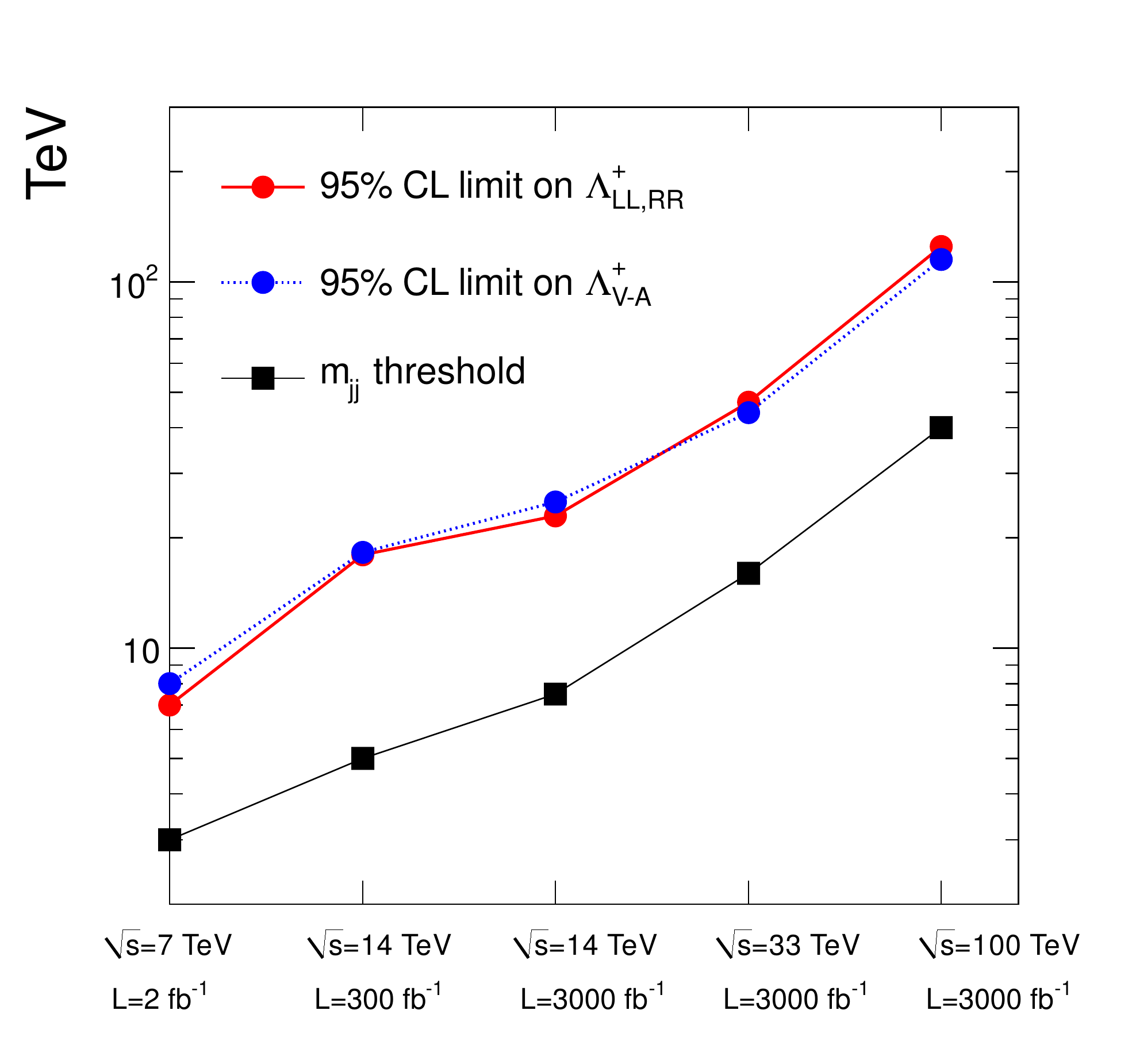}
\caption{Summary of the threshold requirement on $m_{jj}$ and the
  expected limit on composite scale $\Lambda_{LL,RR}$ at each of the
  proposed facilities.}
\label{fig:summary}
\end{figure}

We have studied the sensitivity of future $pp$ collider facilities to
quark compositeness via dijet angular distributions.  As seen in
Fig.~\ref{fig:summary}, increases in center-of-mass energy brings
significant increases in sensitivity to the mass scale, $\Lambda$,
such that a collider with $\sqrt{s}=100$ TeV would be expected to
probe scales above $\Lambda=125$ TeV.

\subsection{Acknowledgements}

We acknowledge useful conversations with LianTao Wang.  DW and SU are
supported by a grant from the Department of Energy Office of
Science. LA and NV are supported by a grant from the National Science
Foundation.  Fermilab is operated by Fermi Research Alliance, LLC
under Contract No. De-AC02-07CH11359 with the United States Department
of Energy.

\bibliography{compos}

\vspace{10cm}

\end{document}